\documentclass[journal=jacsat,manuscript=article]{achemso}

\usepackage[version=3]{mhchem} 
\usepackage{amsmath}
\usepackage{braket}
\usepackage{xcolor}

\author{Yu Wang}
\affiliation{Department of Chemistry, School of Science, Westlake University, Hangzhou 310024 Zhejiang, China}
\altaffiliation
{Institute of Natural Sciences, Westlake Institute for Advanced Study, Hangzhou 310024 Zhejiang, China}
\author{Wenjie Dou}
\email{douwenjie@westlake.edu.cn}
\affiliation{Department of Chemistry, School of Science, Westlake University, Hangzhou 310024 Zhejiang, China}
\altaffiliation
{Institute of Natural Sciences, Westlake Institute for Advanced Study, Hangzhou 310024 Zhejiang, China}

\title
  {Electron transfer at molecule-metal interfaces under Floquet engineering: Rate constant and Floquet Marcus theory}

\begin{document}








\begin{abstract}
Electron transfer (ET) at molecule-metal or molecule-semiconductor interfaces 
is a fundamental  reaction that underlies all electro-chemical and molecular electronic processes 
as well as substrate-mediated surface photochemistry.
In this study, we show that ET rates near a metal surface can be significantly manipulated 
by periodic modulations of an impurity level of the molecule near a metal surface. We employ the analytical Marcus theory and two numerical Floquet surface hopping algorithms that are developed previously, to calculate the ET rates near metal surface as a function of driving amplitudes and driving frequencies. We find that ET rates become faster with increasing the driving amplitude but no turnover effect, while have a turnover effect with increasing driving frequencies.
\end{abstract}

\section{Introduction}
Electron transfer (ET) at molecule-metal or molecule-semiconductor interfaces is of interest to many research fields\cite{lindstrom2006photoinduced},
such as surface photochemistry\cite{zhang2017surface}, heterogeneous catalysis\cite{ye2020nitrogen}, chemisorption\cite{wang2020role}, and dye-sensitized solar cells (DSSCs)\cite{listorti2011electron}. 
There are many approaches
to improve the ET processes which are, for example, 
by using acceptor units with strong electron withdrawing capability within dye molecules for DSSCs\cite{wu2013influence},
or by constructing heterostructures to form a heterojunction in photocatalysis\cite{zhang2017heterostructures}.
Such approaches are all accompanied by chemical structure variations. There is an alternative to modulate ET rates without chemical structure modifications which is the Floquet engineering\cite{phuc2019control}.

Floquet engineering can be served as a powerful tool to systematically understand the
various phenomenon of matter being in light-matter couplings or electromagnetic fields,
through which the non-equilibrium processes are triggered\cite{bukov2015universal,oka2019floquet}.
Under Floquet engineering, the quantum systems are controlled by time-periodic external fields,
which can be used as a theoretical model for matter being manipulated by continuous laser irradiation.
To explore phenomena under external time-periodic drivings,
Floquet theorem has been applied for laser-driven atoms\cite{dundas2000ionization}, strongly correlated electron systems\cite{Bukov2016,bloch2022strongly},
electron-phonon systems\cite{Babadi2017, Hbener2018}, and open transport problems\cite{Tikhonov2002,KOHLER2005}.
Numerous methods are based on the Floquet theory, including Floquet Green function\cite{Wu2008,Martinez2003,Chen2013}, Floquet dynamical mean-field theory (DMFT)\cite{Qin2017,Sandholzer2019},
Floquet scattering theory\cite{Moskalets2002,Li2018,Moskalets2014}, and so forth.
Recently, we developed the Floquet surface hopping (FSH) algorithms from a Floquet classical quantum equation (FCME),
especially for dynamics of an impurity level (a molecular or an atomic orbital) 
under Floquet engineering near a metal surface\cite{wang2023nonadiabaticjcp}.  
The FSH algorithms are valid in conditions of high temperature 
($kT\gg\hbar\omega$, where $\omega$ is the nuclear frequency)
and weak molecule-metal interactions ($kT\gg\Gamma$, where $\Gamma$ is the molecule-metal coupling).
When molecule-metal interactions are stronger than the nuclear oscillations ($\Gamma\gg\hbar\omega$)
so that electronic dynamics are much faster than nuclear motion,
the FCME can be further mapped onto the Floquet Fokker-Planck (FFP) equation.
The FFP equation can be solved by Floquet electronic friction (FEF) Langevin dynamics
that assuming the electron moves on a mean-field potential,
along with electronic friction and random force\cite{wang2023nonadiabatic}.

In this study, we explore the ET rates at molecule-metal interface
as a function of  Floquet drivings.
The model we applied here can be viewed as an extension of spin-boson model with one level of donor coupled to continuous levels of acceptors.
Note that at small molecule-metal couplings $\Gamma$ (nonadaibatic limit) , ET rates can be solved analytically by Marcus theory. 
We find that there is no turnover effect of ET rates as a function of driving amplitudes, while there is a turnover effect of ET rates as a function of driving frequencies. This trends can be explained by analytical gradient of Marcus rate with respect to driving amplitudes and driving frequencies, respectively. Our Flqouet surface hopping algorithms agree well with Marcus theory in small $\Gamma$ regime, and it turns out that in large $\Gamma$ regime, ET rates hold the same trend with respect to the drivings.

\section{Theory}
We consider ET at molecule-metal interface, 
with the molecular energy level coupling to one oscillator and the electronic bath of metal surface.
The Hamiltonian of the total system is given by
\begin{equation}\label{eq0}
    \hat{H} = \hat{H}_S + \hat{H}_B + \hat{H}_T
\end{equation}
\begin{equation}\label{eq1}
    \hat{H}_S = (E(x) + A\sin(\Omega t))d^+d  + V_0(x) + \frac{p^2}{2m}
\end{equation}
\begin{equation}\label{eq2}
    \hat{H}_B = \sum_k \epsilon_k c_k^+ \epsilon_k
\end{equation}
\begin{equation}\label{eq3}
    \hat{H}_T = \sum_k V_k(d^+ c_k + c_k^+ d)
\end{equation}
where $d (d^+)$ and $c_k (c_k^+)$ are the annihilation (creation) operators for 
an electron in the molecular energy level (subsystem)
and in the metal surface (electronic bath),  
$E(x)$ can be an arbitrary function of nuclear position.
If $V_0(x)$ is the diabatic potential energy surface (PES) for the unoccupied state,
we define the diabatic PES for the time-independent occupied state as
$V_1(x)=V_0(x)+E(x)$.
Taking the harmonic potentials approximation,
the parabolic diabatic potential $V_0$ is defined as 
 \begin{equation}\label{v0}
    V_0 (x)= \frac{1}{2}m\omega^2x^2
\end{equation} 
where $\omega$ is the nuclear frequency, and $m$ is the mass of the oscillator.
Herein, we chose $E(x)=gx\sqrt{2m\omega/\hbar}+E_d$,
where $g$ represents electron-phonon (el-ph) coupling strength
and $E_d$ is the energy of the occupied molecular level without Floquet drivings.
In this way,  the parabolic diabatic potential $V_1$ is defined as
\begin{equation}\label{v1}
    V_1 (x)= \frac{1}{2}m\omega^2x^2 + gx\sqrt{2m\omega/\hbar}+E_d
\end{equation} 
The periodic driving acting on the molecular energy level has 
a driving amplitude $A$ and 
a driving frequency $\Omega$ in this model.
The key parameters for this model are the nuclear frequency ($\omega$),
the temperature of the metal surface ($T$), 
the el-ph coupling strength ($g$),
and the molecule-metal interaction ($\Gamma(\epsilon)=2\pi\sum_k|V_k|^2\delta(\epsilon_k-\epsilon)$),
which can be assumed as a constant under the wide band approximation.

In present work, we only focus on the high temperature limit ($kT\gg\hbar\omega,\Gamma$) to explore the electron transfer rates with the influence of different Floquet drivings. 
We employ the Floquet surface hopping algorithm that is proposed in Ref. \cite{wang2023nonadiabaticjcp} to calculate the nonadiabatic dynamics of electronic populations. For each dynamics we can extract ET rate by exponential fitting\cite{Landry2011}.


For Floquet surface hopping (FSH) dynamics,
phase space densities can be propagated in real time.
Let $P_0(x,p,t)$ ($P_1(x,p,t)$) be the probability density 
for the electronic molecular level to be unoccupied (occupied)
at time $t$ with nucleus at position ($x,p$) in phase space.
The Floquet classical master equations (FCME) can be expressed as
\begin{equation}\label{eqcme0}
\begin{split}
        \frac{\partial P_0(x,p,t)}{\partial t}= & \frac{\partial V_0(x,p)}{\partial x} \frac{\partial P_0(x,p,t)}{\partial p} - \frac{p}{m}\frac{\partial P_0(x,p,t)}{\partial x} \\ & - \frac{\Gamma}{\hbar}\Re(\tilde{f}(E(x),t))P_0(x,p,t) + \frac{\Gamma}{\hbar}(1-\Re(\tilde{f}(E(x),t)))P_1(x,p,t)
\end{split}
\end{equation}
\begin{equation}\label{eqcme1}
\begin{split}
    \frac{\partial P_1(x,p,t)}{\partial t} = &\frac{\partial V_1(x,p)}{\partial x} \frac{\partial P_1(x,p,t)}{\partial p} - \frac{p}{m}\frac{\partial P_1(x,p,t)}{\partial x} \\ & + \frac{\Gamma}{\hbar}\Re(\tilde{f}(E(x),t))P_0(x,p,t) - \frac{\Gamma}{\hbar}(1-\Re(\tilde{f}(E(x),t)))P_1(x,p,t)
\end{split}
\end{equation}
Here, $\tilde{f}(E(x),t))$ is a Bessel function modified
Fermi function, which is
\begin{equation}\label{eqfermi}
\begin{split}
    \tilde{f}(E(x),t)) = (\sum_{n,m}(i)^n(-i)^m J_n(z) J_m(z) e^{i(n-m)\Omega t} f(E(x)-m\Omega)
\end{split}
\end{equation}
and $f(E(x)-m\Omega)=1/(1+e^{\beta(E(x)-m\Omega)})$ is the original definition of Fermi function,
where $\beta\equiv1/kT$.
Note that only the real part of $\tilde{f}(E(x),t))$ is employed in FCME.
When taking the cycle average, Eq. (\ref{eqfermi}) becomes
\begin{equation}\label{eqafermi}
\begin{split}
    \bar{\tilde{f}}(E(x)) = \sum_{n} J_n(z)^2 f(E(x)-n\Omega)
\end{split}
\end{equation}

There are two surface hopping algorithms to solve this FCME.
One is propagating electronic density of each trajectory
via FCME directly using equations
\begin{equation}\label{eqp0}
\begin{split}
        \frac{\partial P_0(x,p,t)}{\partial t}= - \frac{\Gamma}{\hbar}\Re(\tilde{f}(E(x),t))P_0(x,p,t) + \frac{\Gamma}{\hbar}(1-\Re(\tilde{f}(E(x),t)))P_1(x,p,t),
\end{split}
\end{equation}
\begin{equation}\label{eqp1}
\begin{split}
    \frac{\partial P_1(x,p,t)}{\partial t} = \frac{\Gamma}{\hbar}\Re(\tilde{f}(E(x),t))P_0(x,p,t) - \frac{\Gamma}{\hbar}(1-\Re(\tilde{f}(E(x),t)))P_1(x,p,t),
\end{split}
\end{equation}
We denote this algorithm as FaSH-density for shorthand notation, which is the same as that in Ref. \cite{wang2023nonadiabaticjcp}.
The other one is solving the FCME through surface hopping between two diabatic potentials $V_0$ and $V_1$, where the hopping rates are cycle-averaged
\begin{equation}\label{eqrate01}
\begin{split}
    \gamma_{0\rightarrow1} = \frac{\Gamma}{\hbar}\bar{\tilde{f}}(E(x)),
\end{split}
\end{equation}
\begin{equation}\label{eqrate10}
\begin{split}
    \gamma_{1\rightarrow0} = \frac{\Gamma}{\hbar}(1-\bar{\tilde{f}}(E(x))).
\end{split}
\end{equation}
We denote this algorithm as the FaSH.


In the limit of small $\Gamma$, the electron transfer rate can be analytically calculated by Marcus theory\cite{Dou2015}.  
\begin{equation}\label{eqk01}
\begin{split}
   k_{0\rightarrow 1}=\int_{-\infty}^{+\infty}d\epsilon\Gamma \bar{\tilde{f}}(\epsilon) \frac{e^{-(E_r-\epsilon+\bar{E}_d)^2}/4E_rkT}{\sqrt{4\pi E_rkT}},
\end{split}
\end{equation}
\begin{equation}\label{eqk10}
\begin{split}
   k_{1\rightarrow 0}=\int_{-\infty}^{+\infty}d\epsilon\Gamma (1-\bar{\tilde{f}}(\epsilon)) \frac{e^{-(E_r+\epsilon-\bar{E}_d)^2}/4E_rkT}{\sqrt{4\pi E_rkT}},
\end{split}
\end{equation}
where $E_r=g^2/\hbar\omega$ is the reorganization energy, 
and $\bar{E}_d=E_d-E_r$ is the renormalized single electron energy,
which also represents the difference in energy between relaxed donor and acceptor (i.e., the "driving force").
In such small $\Gamma$ regime, Marcus nonadiabatic rate is consistent with the surface hopping picture\cite{Dou2015}.

The FEF algorithm is present in the Appendix.

\section{Results}
\begin{figure}
  \centering
  \includegraphics[scale=0.4]{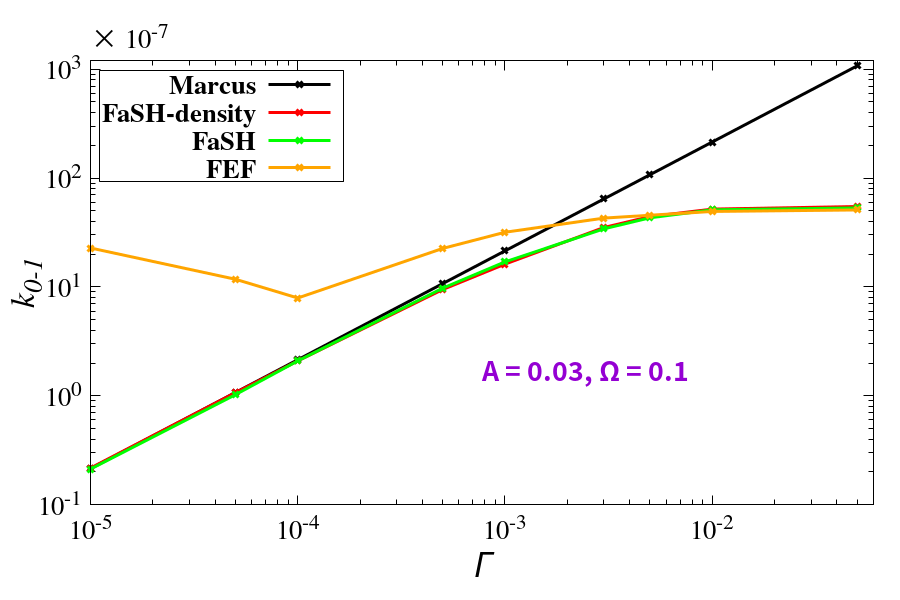}
  \caption{Electron transfer rates near metal surface as a function of $\Gamma$ for four algorithms. $\omega=0.003$, $g=0.025$, $E_d=\frac{g^2}{\hbar\omega}$, $A=0.03$, $\Omega=0.1$, nuclear friction $\gamma_{n}=0.003$. Note that Marcus theory is valid in small $\Gamma$ regime, and FEF method is valid in large $\Gamma$ regime.}
  \label{fig:1}
\end{figure}

It has been revealed preciously that without Floquet driving,
the ET rates have the following trend with increasing the molecule-metal coupling $\Gamma$:
(i) linear in $\Gamma$ that agrees with Marcus theory when $\Gamma$ is small;
(ii) gradually accord with exponential rise as further increasing $\Gamma$ that straying from the Marcus theory\cite{Ouyang2016}.
This is because in small $\Gamma$ limit, 
the probability for an electron to hop is increased with increasing the $\Gamma$,
while in large $\Gamma$ limit,
the rate increases because the adiabatic barrier decreases.
Here, the ET rates from metal surface to molecule ($k_{0\rightarrow 1}$)
as a function of $\Gamma$ under a certain strength of Floquet driving is depicted in Figure \ref{fig:1}.
For FSH and FEF approaches, the external source of nuclear friction $\gamma_n=\hbar\omega$
is taken into account.
We can see that Marcus rate only works well in the small $\Gamma$ limit, while FEF algorithm is only accurate in the large $\Gamma$ limit ($\Gamma > 0.005$), which is consistent with previous conclusion\cite{Ouyang2016}.

Firstly, we set $\Gamma=0.0005$ where the Marcus nonadiabatic theory works, 
to investigate variation trends of rates as a function of Floquet driving strengths.

\begin{figure}
  \centering
  \includegraphics[scale=0.13]{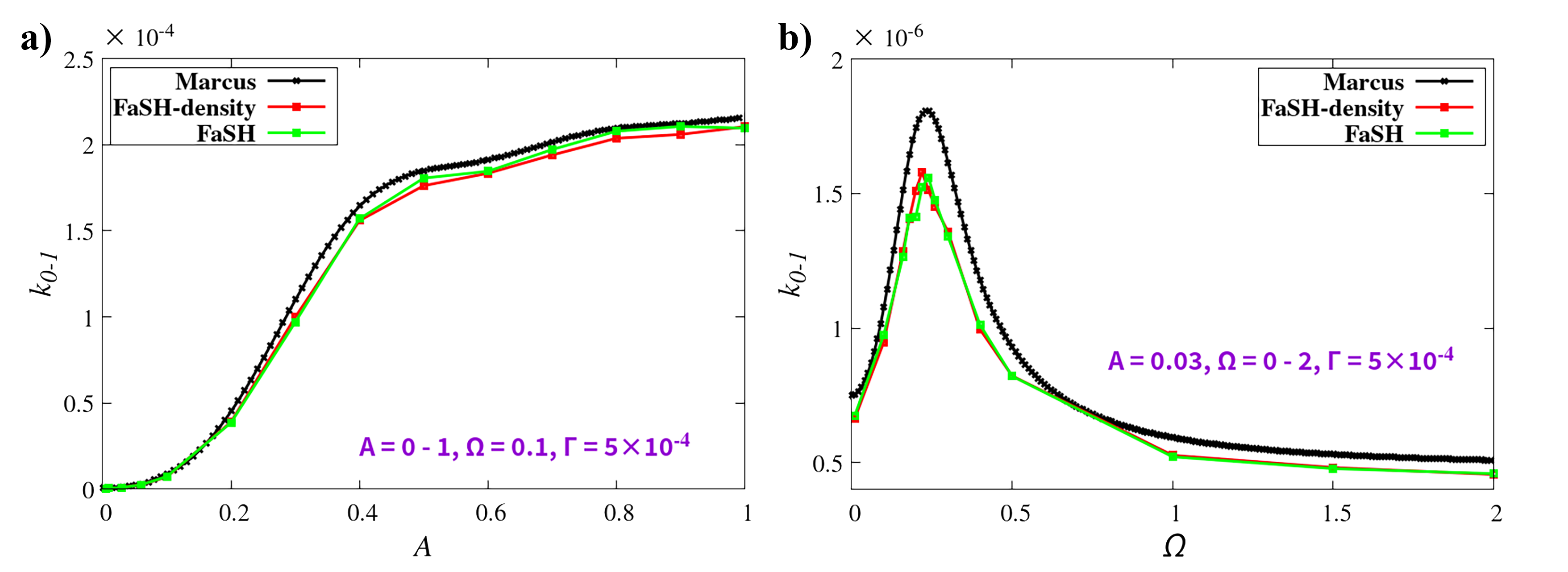}
  \caption{Electron transfer rates near metal surface as the function of a) driving amplitudes $A$ ($\Omega=0.1$); b) driving frequencies $\Omega$ ($A=0.03$) at small $\Gamma$ limit ($\Gamma=0.0005$). $\omega=0.003$, $g=0.025$, $E_d=\frac{g^2}{\hbar\omega}$. Note that numerical surface hopping algorithms are consistent with Marcus rates in small $\Gamma$ limit. There is no turnover effect of ET rates as a function of $A$, while there is a turnover effect as a function of $\Omega$.}
  \label{fig:2}
\end{figure}

\begin{figure}
  \centering
  \includegraphics[scale=0.13]{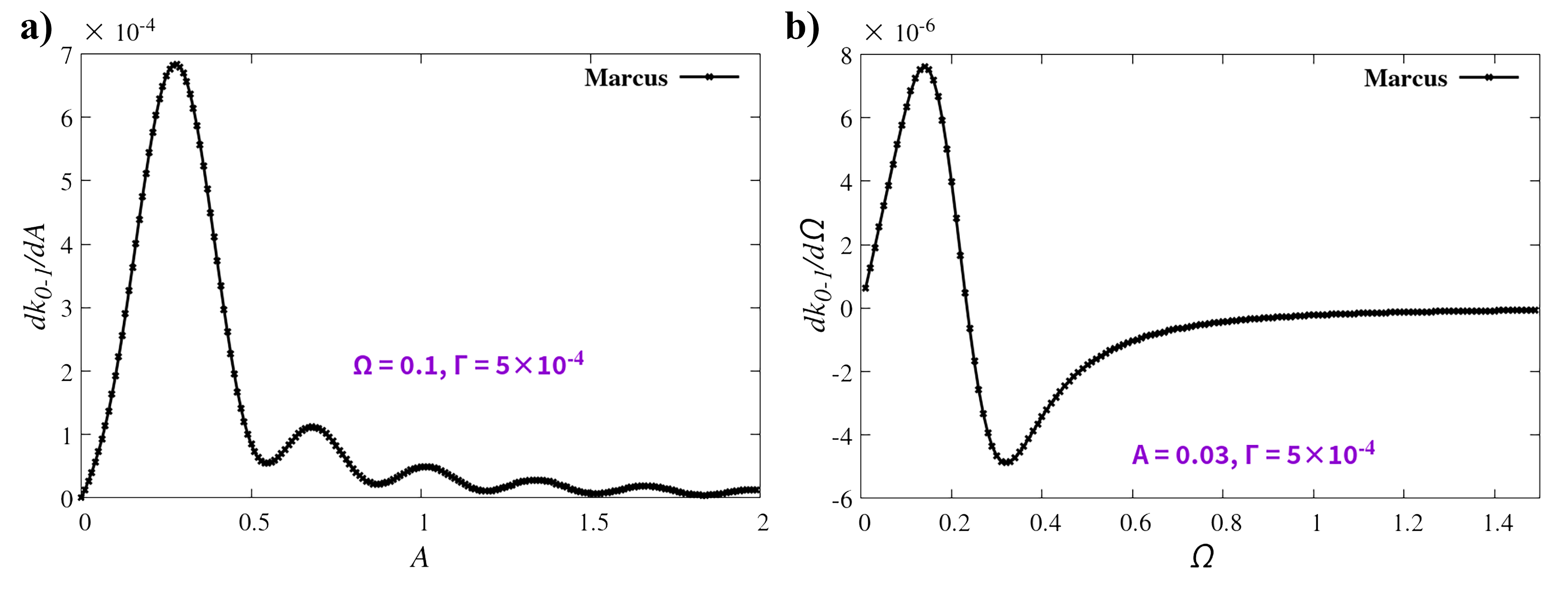}
  \caption{Analytical gradient of $k_{0\rightarrow 1}$ with respective to a) driving amplitude $A$ and b) driving frequency $\Omega$. $\omega=0.003$, $g=0.025$, $E_d=\frac{g^2}{\hbar\omega}$.  Note that $k_{0\rightarrow 1}/dA \geq 0$ which means $k_{0\rightarrow 1}$ increases with $A$ until reaches to a plateau. While $k_{0\rightarrow 1}/d\Omega$ cross over zero point, thus demonstrating the turnover effect.}
  \label{fig:3}
\end{figure}

The ET rates as a function of driving amplitudes $A$ are
shown in Figure \ref{fig:2}a. Here, the driving frequency is fixed at $\Omega=0.1$.
We can see that the results of analytical (Marcus) and numerical methods (FaSH-density and FaSH) are consistent in such small $\Gamma$ limit. 
The ET rates increase rapidly with the driving amplitude $A$ at the beginning,
and then gradually level off.
Notice that there is no turnover effect of rates as increasing driving amplitudes 
at molecule-metal surface, due mainly to the continuous levels in metal surface 
so that there is no Marcus inverted region.
To prove this no turnover effect, we further take the analytical gradient of $k_{0\rightarrow 1}$ with respect to $A$ ($\frac{dk_{0\rightarrow 1}}{dA}$), which is shown in Figure \ref{fig:3}a. We can see that no gradient value is below zero which verifying ET rates as a function of $A$ on the metal surface has no turnover effect..  

The ET rates as a function of driving frequencies $\Omega$ are
shown in Figure \ref{fig:2}b. Here, the driving amplitude is fixed at $A=0.03$.
Note that there is a turn over effect as increasing the $\Omega$. The analytical gradient of $k_{0\rightarrow 1}$ with respect to $\Omega$ ($\frac{dk_{0\rightarrow 1}}{d\Omega}$) is shown in Figure \ref{fig:3}b. Note that the gradient cross over the zero point which verifying the turnover effect of ET rates. When keep increasing the driving frequencies, the ET rates reach to a plateau
as if there is no drivings.

\begin{figure}
  \centering
  \includegraphics[scale=0.13]{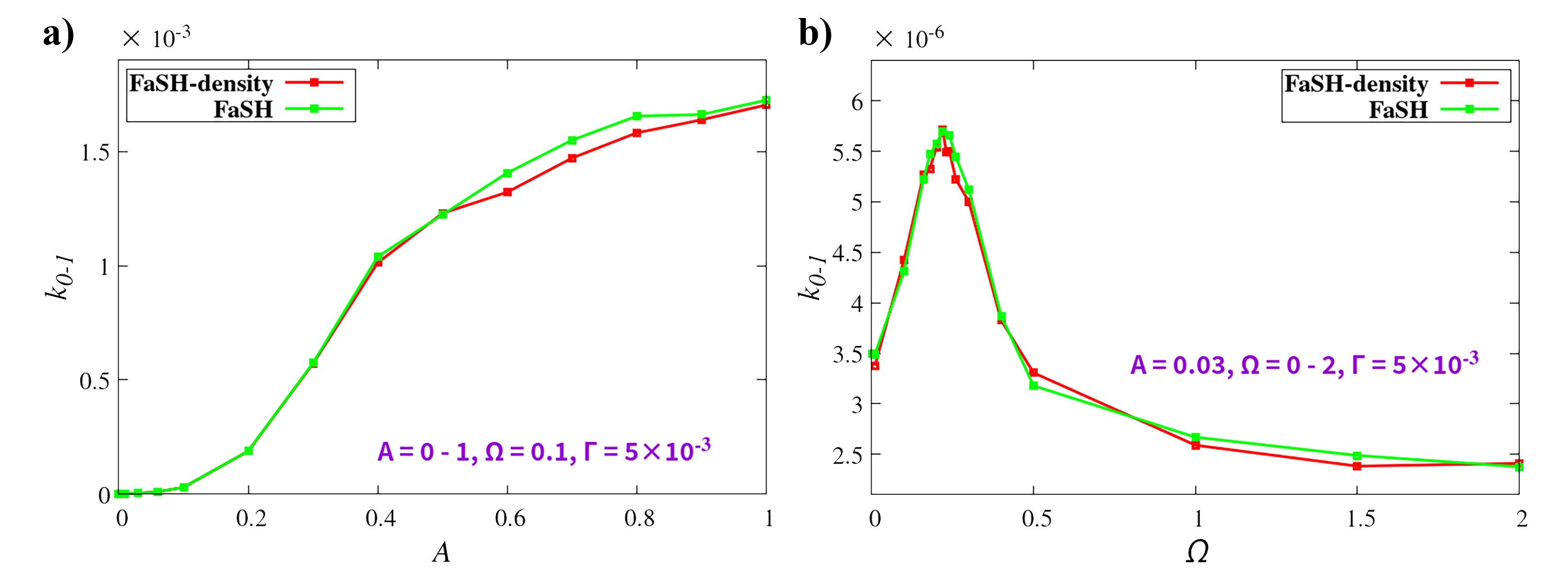}
  \caption{Electron transfer rates near metal surface as the function of a) driving amplitudes $A$ ($\Omega=0.1$); b) driving frequencies $\Omega$ ($A=0.03$) at large $\Gamma$ limit ($\Gamma=0.005$). $\omega=0.003$, $g=0.025$, $E_d=\frac{g^2}{\hbar\omega}$. We see the same trends of ET rates as the function of $A$ and $\Omega$ with that in Figure \ref{fig:2}.}
  \label{fig:4}
\end{figure}

Following, we increase the molecule-metal interaction as $\Gamma=0.005$, in which regime Marcus rate fails as shown in Figure \ref{fig:1}.
The ET rates as a function of $A$ and $\Omega$ predicted by numerical surface hopping algorithms have the same trend with that when $\Gamma=0.0005$, which can be seen from Figure \ref{fig:4}. Notice that when $\Gamma$ is increased, ET rates are also increased as a whole.

\begin{figure}
  \centering
  \includegraphics[scale=0.4]{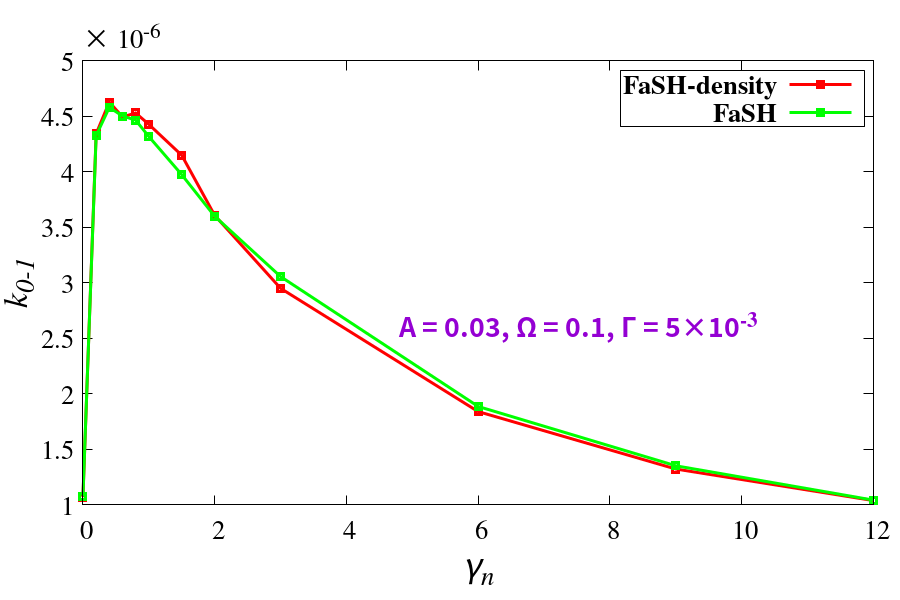}
  \caption{Electron transfer rates near metal surface as a function of phonon friction $\gamma_n$ under a Floquet driving ($A=0.03, \Omega=0.1$) at large $\Gamma$ limit. $\omega=0.003$, $g=0.025$, $E_d=\frac{g^2}{\hbar\omega}$.}
  \label{fig:5}
\end{figure}

Additionally, we simulate the ET rates as a function of nuclear frictions $\gamma_n$ in Figure \ref{fig:5} under certain external driving.
The turnover phenomenon under Floquet engineerings is just the same with
that without any periodic drivings\cite{Dou2015friction}.

\section{Conclusion}
To sum up, we calculate ET rates near metal surface as a function of periodic driving amplitudes and driving frequencies. When molecule-metal coupling is small, we compare ET rates of analytical Marcus rate with numerical Floquet surface hopping algorithms. We find that ET rates have no turnover effect as a function of driving amplitudes, while have a turnover effect as a function of driving frequencies. 
When molecule-metal coupling is larger, the ET rates have the same trend with different drivings.
We also demonstrate that system with Floquet engineerings also show a turnover effect of ET rates
as a function of nuclear friction damping.
Given the ET rates at molecule-metal can be
significantly manipulated via Floquet engineerings,
it would be applied extensively in domains of photocatalysis,
dye-sensitized solar cells, chemisorptions and so forth.

\section{Appendix}
\subsection{Floquet Electronic Friction (FEF) algorithm}

The nuclear motion in FEF is Ref. \cite{wang2023nonadiabatic},
\begin{equation}\label{eqLD1}
\begin{split}
   \dot{p} = - \frac{\partial U}{\partial x} - \gamma_ep + \xi ,
\end{split}
\end{equation}
\begin{equation}\label{eqLD2}
\begin{split}
   \dot{x} =  \frac{p}{m}
\end{split}
\end{equation}
where $\xi$ is the random force that is assumed to be a
Gaussian variable with a norm $\sigma=\sqrt{2m\gamma_e'kT/dt}$ ($dt$ is the time step interval).
The correlation function of the random force is $D(x)=\gamma_e'(x) mkT$,
where, $\gamma_e'(x)$ is,
\begin{equation}\label{eqgamma_e'}
\begin{split}
   \gamma_e'(x) = \frac{\beta}{\Gamma M}\left( \frac{dE(x)}{dx} \right)^2   \bar{\tilde{f}} (1-\bar{\tilde{f}} )
\end{split}
\end{equation}
Here, $U$ is time-averaged potential of mean force,
\begin{equation}\label{eq_a_U}
\begin{split}
   \bar{U}(x) = \frac{1}{2}\hbar\omega x^2 - \frac{1}{\beta}\sum_{n}|J_n(z)|^2 log(1+exp(-\beta(E(x)-n\Omega))
\end{split}
\end{equation}
and $\gamma_e$ is the electronic friction,
\begin{equation}\label{eqgamma_e}
\begin{split}
   \gamma_e(x) & =  -\frac{1}{\Gamma M}\frac{dE(x)}{dx}
   \frac{\partial \bar{\tilde{f}} }{\partial x}
\end{split}
\end{equation}

\subsection{Gradient of Marcus rate}

\begin{equation}
\begin{split}
   \frac{\partial k_{0\rightarrow 1}}{\partial A} & = \int_{-\infty}^{+\infty}d\epsilon\Gamma  \frac{e^{-(E_r-\epsilon+\bar{E}_d)^2}/4E_rkT}{\sqrt{4\pi E_rkT}} \times  
   \sum_{n} \left[\frac{1}{\Omega}J_{n}(\frac{A}{\Omega})(J_{n-1}(\frac{A}{\Omega})-J_{n+1}(\frac{A}{\Omega}))f(\epsilon-n\Omega)\right]
\end{split}
\end{equation}
\begin{equation}
\begin{split}
   \frac{\partial k_{0\rightarrow 1}}{\partial \Omega} & = \int_{-\infty}^{+\infty}d\epsilon\Gamma  \frac{e^{-(E_r-\epsilon+\bar{E}_d)^2}/4E_rkT}{\sqrt{4\pi E_rkT}} \times  \\ & 
   \sum_{n} \left[-\frac{A}{\Omega^2}J_{n}(\frac{A}{\Omega})(J_{n-1}(\frac{A}{\Omega})-J_{n+1}(\frac{A}{\Omega}))f(\epsilon-n\Omega)
   + \frac{n}{kT}J_n(\frac{A}{\Omega})^2 f(\epsilon-n\Omega)(1-f(\epsilon-n\Omega))\right]
\end{split}
\end{equation}

\begin{acknowledgement}

This material is based upon work supported by National Science Foundation of China (NSFC No. 22273075)

\end{acknowledgement}




\bibliography{achemso-demo}

\end{document}